\documentclass{article}

\usepackage{arxiv}

\usepackage[utf8]{inputenc} 
\usepackage[T1]{fontenc}    
\usepackage{hyperref}       
\usepackage{url}            
\usepackage{booktabs}       
\usepackage{amsfonts}       
\usepackage{nicefrac}       
\usepackage{microtype}      
\usepackage{lipsum}

\title{Generative Synthesis of Insurance Datasets}

\author{
    Kevin Kuo
   \\
    Kasa AI \\
   \\
  \texttt{\href{mailto:kevin@kasa.ai}{\nolinkurl{kevin@kasa.ai}}} \\
  }

\usepackage{makecell}
\usepackage{amsmath}
\usepackage[numbers]{natbib}
\setcitestyle{authoryear,open={(},close={)}}
\usepackage{graphicx}
\usepackage{mathtools}
\usepackage{lmodern}

\usepackage{booktabs}
\usepackage{longtable}
\usepackage{array}
\usepackage{multirow}
\usepackage{wrapfig}
\usepackage{float}
\usepackage{colortbl}
\usepackage{pdflscape}
\usepackage{tabu}
\usepackage{threeparttable}
\usepackage{threeparttablex}
\usepackage[normalem]{ulem}
\usepackage{makecell}
\usepackage{xcolor}

\begin{document}
\maketitle

\def\tightlist{}

\begin{abstract}
One of the impediments in advancing actuarial research and developing open
source assets for insurance analytics is the lack of realistic publicly
available datasets. In this work, we develop a workflow for synthesizing
insurance datasets leveraging CTGAN, a recently proposed neural network
architecture for generating tabular data. Applying the proposed workflow to
publicly available data in the domains of general insurance pricing and life
insurance shock lapse modeling, we evaluate the synthesized datasets from a
few perspectives: machine learning efficacy, distributions of variables, and
stability of model parameters. This workflow is
implemented via an R interface to promote adoption by researchers and data
owners.
\end{abstract}


\hypertarget{introduction}{%
\section{Introduction}\label{introduction}}

With the increased interest in applying machine learning (ML) and predictive modeling
techniques across all fields of actuarial science in recent years, access to data is becoming more
important. Having access to realistic data from insurers allows researchers to
tackle more practical problems and validate newly developed methodologies. If the
data is also publicly available, it allows researchers and companies to open source
their methodologies, which encourages others to build upon existing work. Furthermore,
having a common collection of datasets for each research question allows the community to
define the state of the art, benchmark new workflows, and measure progress.

Of course, much of the data in the industry is confidential and proprietary. While
there are publicly available datasets, new datasets are hard to come by. Even if
data owners are willing
to share anonymized datasets, the effort involved in obsfucating the data and
navigating beauracracy may prevent them from doing so. We posit that, if there is
an easy way for data owners to create ``fake'', or synthesized, data that have
characteristics similar to the real data, it would reduce friction in data disclosure.

While synthetic data generation is an active area of research in the broader ML
community, with many recent results (see, e.g., \citet{choi2017generating}, \citet{Park_2018}, and
\citet{xuModelingTabular2019}), research on synthesizing insurance datasets
in particular
is scarce. One notable example is \citet{gabrielliIndividualClaims2018}, which
describes a methodology for fitting neural networks to claims history data.
The authors provide a fitted model for researchers to generate data, and the model has
been implemented as an R package.\footnote{\url{https://github.com/kasaai/simulationmachine}} However, it does not provide an easy way for a
data owner to develop a new data generator from a different portfolio of claims.

In this paper, we propose a workflow to train a neural network-based data synthesizer
using confidential data and generate data from the trained synthesizer. We utilize the CTGAN
architecture proposed by \citet{xuModelingTabular2019}, which is based on generative
adversarial networks (GAN) \citep{NIPS2014_5423}, and introduce modifications along with
pre- and post-processing transformations specific to insurance datasets. We introduce
an extension of the ML efficacy evaluation methodology from the CTGAN paper
utilizing cross-validation, and evaluate our
workflow on two publicly available datasets using this methodology. To promote
adoption, we provide an R package and code templates for researchers and data owners to use.

The remainder of the paper is organized as follows. Section \ref{methodology} provide
brief overviews of GAN and CTGAN and introduces our workflow, Section \ref{applications}
applies the workflow to two publicly
available datasets, describes our evaluation methodology for ML efficacy, and evaluates
the synthesized datasets with it, Section \ref{workflow} discusses the data disclosure
workflow and data privacy considerations, and Section \ref{conclusion} concludes.

\hypertarget{methodology}{%
\section{Methodology}\label{methodology}}

Our workflow is based on the CTGAN architecture with some modifications. As of
the writing of this paper, CTGAN represents the
state-of-the-art for synthesizing tabular data \citep{xuModelingTabular2019}.
In this section, we provide a cursory overview of GANs, the extensions of GAN that
CTGAN proposes in order to train
quality generative models for tabular data, and modifications we introduce to
adopt the architecture for insurance datasets. We remark that neural network rudiments
have been discussed extensively in the actuarial literature (see, for example, the survey by
\citet{richman2018ai}, the lecture notes by \citet{wuthrich2019data}, and references therein)
and will not be covered here.

\hypertarget{gan-overview}{%
\subsection{GAN Overview}\label{gan-overview}}

\begin{figure}

{\centering \includegraphics[width=0.75\linewidth]{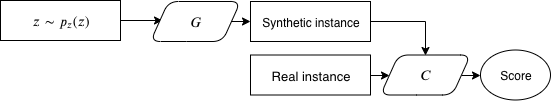} 

}

\caption{Architecture of a GAN.}\label{fig:gan-fig}
\end{figure}

GAN is generative modeling technique based on neural networks. In a typical
setup, a generator network, \(G(Z)\), take some noise vector (usually sampled from
a spherical Gaussian) as input, and outputs an instance of interest (which, in our
case, is a row of data). A discriminator, or critic, network, \(C(X)\)
takes an instance (again, a row in our case) and outputs a scalar score that
represents how likely it thinks the instance comes from the real data distribution.
The generator and critic then play a ``game'' in which the former tries to create more
realistic instances while the latter tries to identify the ``fake'' instances, much
like the relationship between an art forger and a curator, where the instances of
interest are paintings. In fact, in its original formulation, GANs are a minimax game in
the formal sense, and one can write

\begin{equation}\label{eq:gan}
    \min_{G}\max_{C} V(C, G) = \mathbb{E}_{x\sim p_{\text{data}}(x)}[\log C(x)] +
    \mathbb{E}_{z \sim p_z(z)}[\log (1-C(G(z)))],
\end{equation}

where \(V(C, G)\) denotes the value function. Figure \ref{fig:gan-fig} illustrates
the relationship between the generator and the critic.

Again, in Equation \eqref{eq:gan}, the critic \(C\) aims to maximize the score it assigns to
instances from the real data samples, while the generator \(G\) maximize the score
given by the critic to the samples it generates. In practice, we alternately optimize
the two networks, fixing one while training the other. Once the generator and
critic networks have been trained, one can input samples of
\(Z\) into the generator \(G\) to obtain synthesized instances. For more details on GAN,
we refer the user to \citet{NIPS2014_5423}.

\hypertarget{ctgan}{%
\subsection{CTGAN}\label{ctgan}}

\begin{figure}

{\centering \includegraphics[width=0.75\linewidth]{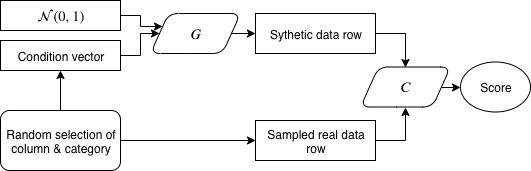} 

}

\caption{Architecture of CTGAN.}\label{fig:ctgan-fig}
\end{figure}

CTGAN, whose architecture we illustrate in Figure \ref{fig:ctgan-fig}, proposes two novel techniques to improve tabular data generation: mode-specific
normalization and conditional training-by-sampling.

Mode-specific normalization
is designed to address the difficulty vanilla GAN has with modeling multi-modal
distributions in numeric columns. It uses variational Gaussian mixture (VGM) modeling
\citep{bishop2006pattern} to
determine the number of modes for each column and normalize the values accordingly.
During training, these encoded values are used
in place of the original data; upon obtaining a synthesized dataset, the values
are transformed back to the original scale.

To address imbalanced factor level frequencies in categorical columns, CTGAN employs
a conditional training approach. In this scheme, the random selection of a column
and one of its levels is encoded into a condition vector. This condition vector
is used both as an input to the generator and a filtering condition for sampling
from the real data distribution. To ensure that rare categorical levels are sampled
evenly, the authors choose to sample according to the log-frequency of the categories;
in other words, the frequency of each category is logged and then normalized.

In addition to these extensions, CTGAN also leverages recent advances in GAN
training such as Wasserstein GAN with gradient penalty \citep{gulrajani2017improved}
and PacGAN \citep{lin2017pacgan}, which improve learning stability and
help avoid mode collapse, a scenario where the generator repeatedly generates
similar instances without diversity.

\hypertarget{modifications-for-insurance-datasets}{%
\subsection{Modifications for Insurance Datasets}\label{modifications-for-insurance-datasets}}

We make one key modification to the CTGAN implementation to accommodate our domain
specific datasets. During sampling of categories, we use the true data frequency
rather than the log-frequency. We find that using log-frequency causes the
synthesizer to unrealistically oversample rare categories.

To complete our workflow, we perform dataset-specific transformations before
training the synthesizer and after the generated dataset is obtained. For
insurance datasets, these include ensuring that numerical columns with few
unique values are
properly learned and that events and exposures are internally consistent. In the
next section, we provide specific examples of these transformations.

\hypertarget{applications}{%
\section{Application Examples}\label{applications}}

We evaluate our methodology on two publicly available datasets: the first is the
French Third-Party Liability (TPL) claim frequency dataset, which is a well
studied pricing dataset first introduced in \citet{charpentierComputationalActuarial2014};
the second is a dataset for life insurance shock lapse modeling originally
provided as part of a Society of Actuaries (SOA) experience study \citep{soa_2014}. Both of these
datasets are available for download online.\footnote{\url{https://cellar.kasa.ai}} In the remainder of this
section, we describe in detail our evaluation methodology, apply it to the two
datasets, and discuss the results.

\hypertarget{evaluation-of-predictive-modeling-efficacy}{%
\subsection{Evaluation of Predictive Modeling Efficacy}\label{evaluation-of-predictive-modeling-efficacy}}

\begin{figure}

{\centering \includegraphics[width=0.75\linewidth]{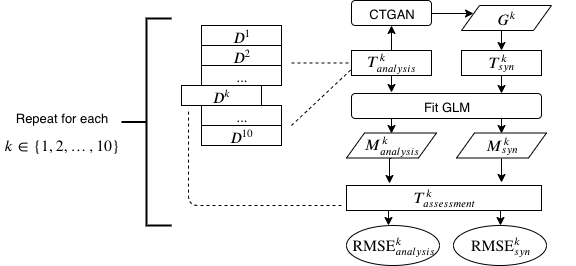} 

}

\caption{ML efficacy validation scheme.}\label{fig:cv-flow}
\end{figure}

Our evaluation scheme extends the ML
efficacy definition, described in \citet{xuModelingTabular2019} and first
introduced by \citet{esteban2017realvalued}, by introducing
a cross-validation component. The process, illustrated in Figure \ref{fig:cv-flow},
is as follows. First, we set up 10-fold cross validation on the modeling
dataset, which we denote as \(D\); in other words, we randomly split the dataset
into 10 subsets of approximately equal size, which are referred to as folds
\(D^k\), \(k = 1, \dots, 10\). We define each \(T_{assessment}^k \coloneqq D^k\) as the
assessment set, and the complement \(T_{analysis}^k = D\setminus D^k\) as
the analysis set. For each \(k\), we train a generative model \(G^k\) using \(T_{analysis}^k\),
and use it to generate a table \(T_{syn}^k\), which has the same number of records as \(T_{analysis}^k\).
Then, we train two predictive models on \(T_{syn}^k\) and \(T_{analysis}^k\),
obtaining models \(M_{syn}^k\) and \(M_{analysis}^k\), respectively. Each of the predictive models is then
used to score \(D^k\) to obtain an out-of-sample performance metric. Finally, we
inspect these metrics to see how much worse the model trained from the
synthetic data performs. For both examples, we use the default hyperparameters
specified in the CTGAN paper. The code for the experiments in this section is
available on GitHub.\footnote{\url{https://github.com/kasaai/gen-syn}}

\hypertarget{tpl-claim-frequency-modeling}{%
\subsection{TPL Claim Frequency Modeling}\label{tpl-claim-frequency-modeling}}

The TPL dataset consists of 678,013 records, each representing a motor insurance
policy in a single year. The dataset contains various policy characteristics,
exposures, and the claim counts associated with each policy. The modeling
problem we consider is predicting the number of claims filed on each policy, i.e.,
the frequency component of a frequency-severity model.
Prior to cross-validation, we employ pre-processing steps as outlined in
\citet{nollCaseStudy2018} to address data quality issues and perform binning of
variables. The variables and their transformations are listed in Table
\ref{tab:frtplvars}.

\begin{table}[!h]

\caption{\label{tab:frtplvars}Columns of the TPL dataset and their transformations.}
\centering
\begin{tabular}[t]{llll}
\toprule
Column & Original type & Pre-processing & Post-processing\\
\midrule
Claim count & Integer & \makecell[l]{Upper bound by four,\\convert to categorical} & Convert to Integer\\
Vehicle power & Categorical & Bin & ---\\
Vehicle age & Numeric & Bin & ---\\
Bonus/malus & Numeric & Bin & ---\\
Vehicle brand & Categorical & --- & ---\\
Vehicle gas & Categorical & --- & ---\\
Density & Numeric & Log & ---\\
Exposure & Numeric & Upper bound by one & ---\\
\bottomrule
\end{tabular}
\end{table}

During training of the synthesizer, we treat the claim count variable, which
takes five different values (0, 1, 2, 3, and 4) as a categorical variable. We
find that this treatment allows the synthesizer to learn a more realistic
distribution. When we treat the variable as numeric, the synthesized samples
underrepresent policies with more than one claim.

In order to constrain training time, we take a subset of the analysis data in
each fold by randomly sampling 100,000 rows. This results in training time of
approximately 10 minutes, which we feel is reasonable for users to experiment
with. During training, we use a minibatch size of 10,000.

After obtaining generated data from the trained synthesizer, we lower bound the
exposure at one day (\(1/365\)), since the synthesizer may generate values less than zero.

A generalized linear model (GLM) is fit to the
training data in each of the cross validation splits. The error distribution
assumed is Poisson with the canonical log link function, and log policy
exposures are used as the offset term. Root mean squared error (RMSE) is used as
the performance metric for comparing models. We remark that no variable
selection or penalized regression techniques are employed, as our purpose is not
to fine-tune a predictive model.

\hypertarget{shock-lapse-modeling}{%
\subsection{Shock Lapse Modeling}\label{shock-lapse-modeling}}

For the other example, we employ the data released with the SOA 2014 Post Level
Term Lapse \& Mortality Report \citep{soa_2014}. This dataset contains aggregated experience from
various companies for term life products from 2000 to 2012. Each
row in the dataset represents a unique combination of study year, cell, and duration.
Here, each cell represents a unique combination of policy and policyholder
characteristics, including gender, issue age, and face amount, among others.

In level premium term life insurance, policyholders pay fixed premiums for a
period of time (ten years in our case study) agreed upon at contract inception.
At the end of the level term, these contracts usually convert to annually renewable
term (ART), which means the premiums increase, typically significantly, year after
year. The increase in premiums results in a phenomenon known as shock lapse, wherein
insurers experience a higher lapse rate than during the level term, leading to
decreased cashflows and a shift towards higher risk policyholders. Being able
to forecast which policyholders are likely to lapse aids with risk management and
assumptions setting at
insurers, which motivates this example. The predictive modeling problem we consider
involves predicting the number of policies that lapse at durations after the
level term, using policy characteristics as predictors.

Following \citet{soa_2014}, we perform further grouping of some categorical variables
and convert some ordinal categorical variables, such as issue age and premium
jump ratio, to numeric. The full list of predictors and their pre-processing
transformations can be found in Table \ref{tab:lapsevars}

\begin{table}[!h]

\caption{\label{tab:lapsevars}Columns of the Lapse Study dataset and their transformations.}
\centering
\begin{tabular}[t]{llll}
\toprule
Column & Original type & Pre-processing & Post-processing\\
\midrule
Lapse count & Integer & Divide by exposure & \makecell[l]{Bound to [0, 1], \\multiply by exposure,\\and round}\\
Risk class & Categorical & Bin & ---\\
Face amount & Categorical & --- & ---\\
Issue age & Catgorical & Convert to numeric & ---\\
Premium jump ratio & Categorical & Convert to numeric & ---\\
Duration & Categorical & --- & ---\\
Exposure & Integer & --- & Lower bound by one\\
\bottomrule
\end{tabular}
\end{table}

Recall that each row of the dataset represents a cell rather than an individual
policy, which means they have a wide range of exposures and lapse counts. Also,
each exposure can lapse at most once, which differs from the previous TPL case
study. To accommodate these specifics, we introduce the following scheme for
data generation. Prior to fitting the synthesizer, we compute the lapse rate,
defined as the number of lapses divided by the number of policies, and use it
in place of the actual lapse count. Once we have the synthesized dataset, we
bound the lapse rate column below by zero and above by one. The lapse rate is then
multiplied by the generated exposures (number of policies) and rounded to obtain
an integer lapse count.

Similar to the TPL example, we use a Poisson GLM, and the (log) number of policies
is used in the offset. RMSE is used as the performance metric.

\hypertarget{results-discussion}{%
\subsection{Results \& Discussion}\label{results-discussion}}

In this section, we present and discuss the results of our experiments. The main findings are
as follows:

\begin{enumerate}
\def\labelenumi{\arabic{enumi}.}
\tightlist
\item
  In terms of cross-validation performance, the models trained using the synthetic
  datasets perform similarly to those trained using the real datasets;
\item
  The distribution of variables, both continuous and categorical, in the synthetic
  datasets are consistent with those in the real datasets; however,
\item
  The relativities of GLM models built on the synthetic datasets \emph{do not} exhibit
  similar patterns to those of models built on real datasets.
\end{enumerate}

\begin{figure}
\centering
\includegraphics{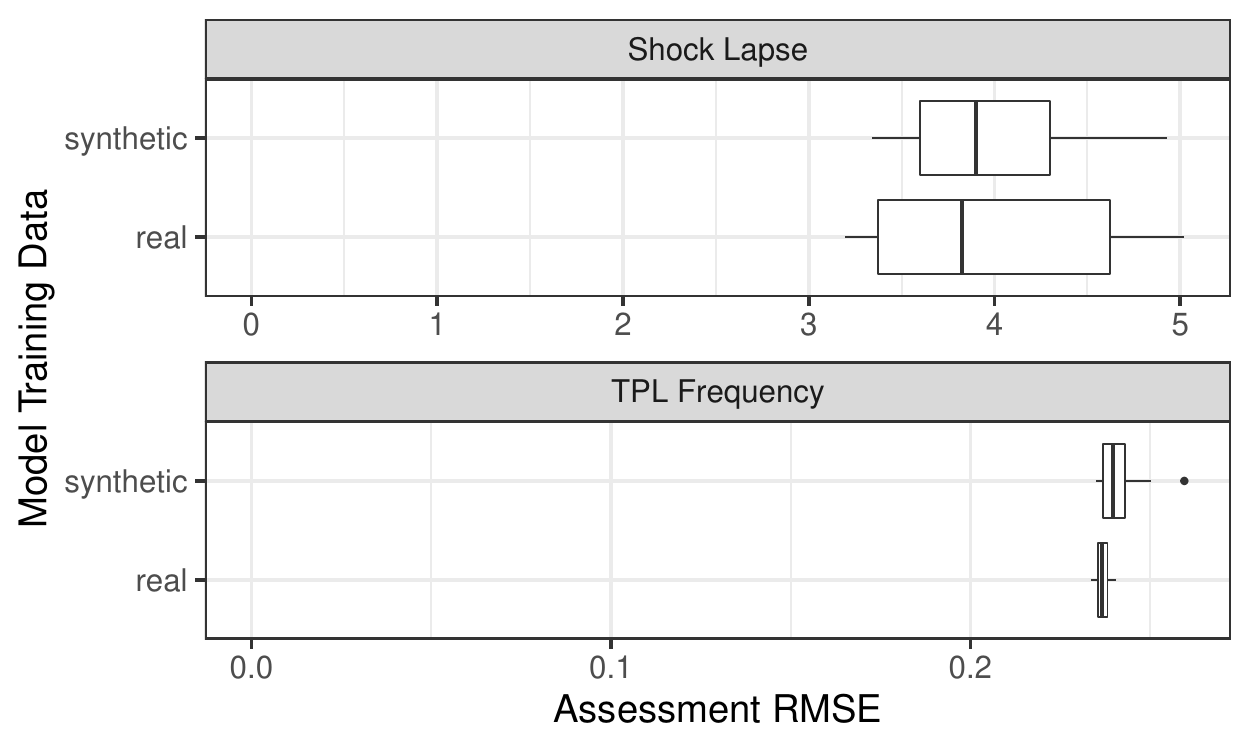}
\caption{\label{fig:metrics}Box plots of cross-validated performance of models trained on real and synthetic data for the two case studies.}
\end{figure}

In Figure \ref{fig:metrics} and Table \ref{tab:metrics-table}, we exhibit the cross-validation performance of models built on
synthetic and real data for the two case studies. We see that the models trained
using the synthetic datasets perform similarly to those trained using the real
datasets.

\begin{table}[!h]

\caption{\label{tab:metrics-table}Average cross-validated metrics of models trained on real and synthetic data.}
\centering
\begin{tabular}[t]{lrrl}
\toprule
Dataset & Mean RMSE (Real Data) & Mean RMSE (Synthetic Data) & Relative Difference\\
\midrule
TPL Frequency & 0.2367 & 0.2419 & 2.21\%\\
Shock Lapse & 4.0038 & 4.0203 & 0.41\%\\
\bottomrule
\end{tabular}
\end{table}

\begin{figure}
\centering
\includegraphics{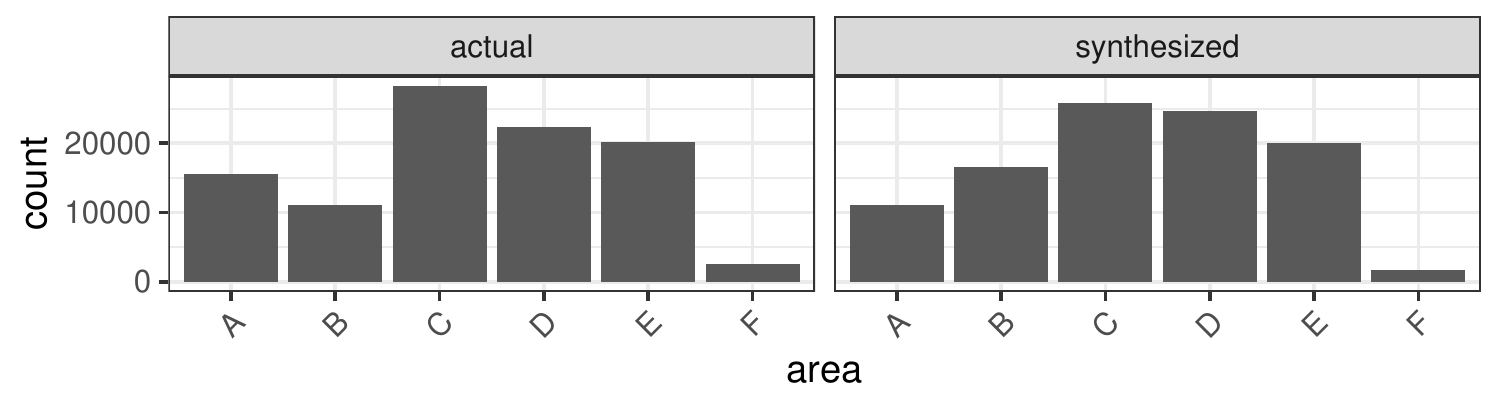}
\caption{\label{fig:distribution-area}Distributions of the Area variable.}
\end{figure}

\begin{figure}
\centering
\includegraphics{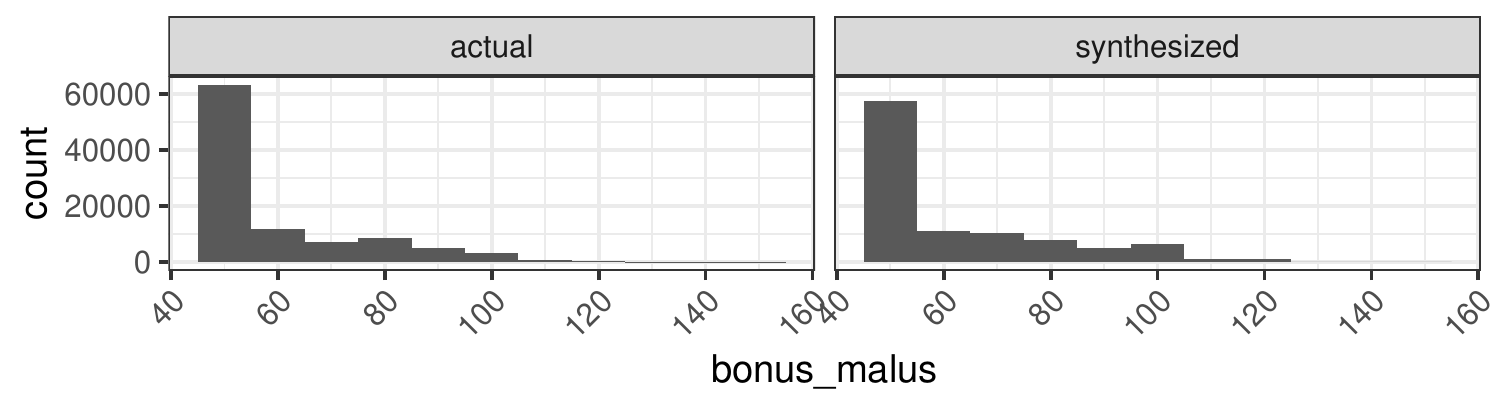}
\caption{\label{fig:distribution-bonus-malus}Distributions of the Bonus/Malus variable.}
\end{figure}

\begin{figure}
\centering
\includegraphics{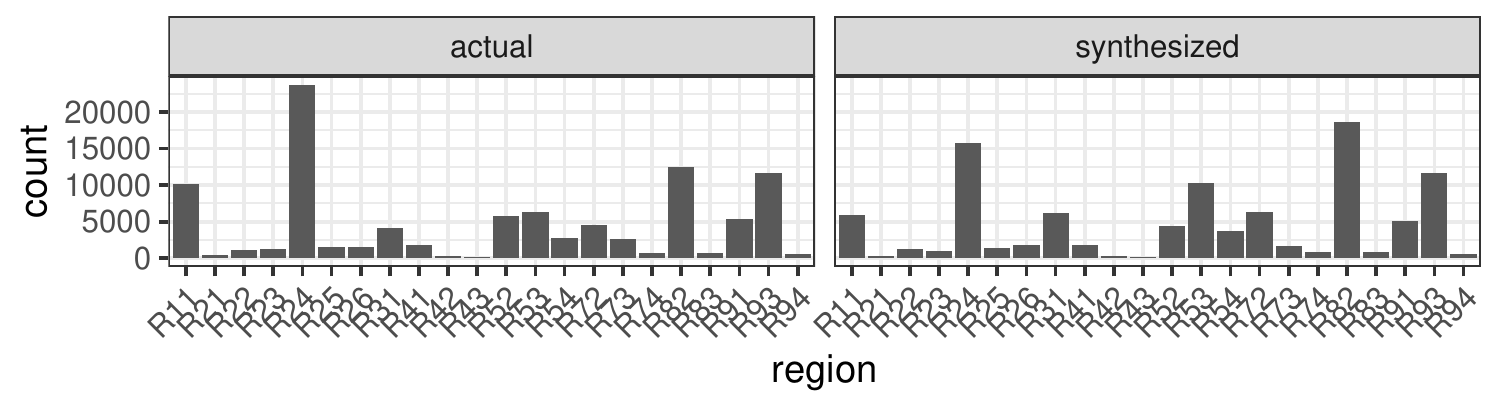}
\caption{\label{fig:distribution-region}Distributions of the Region variable.}
\end{figure}

In Figures \ref{fig:distribution-area}, \ref{fig:distribution-bonus-malus}, and
\ref{fig:distribution-region}, we compare distributions of select variables from
a synthesized sample of the TPL datasets with their distributions from the real
dataset. We note that, qualitatively, the synthesized distributions exhibit realistic
patterns.

\begin{figure}
\centering
\includegraphics{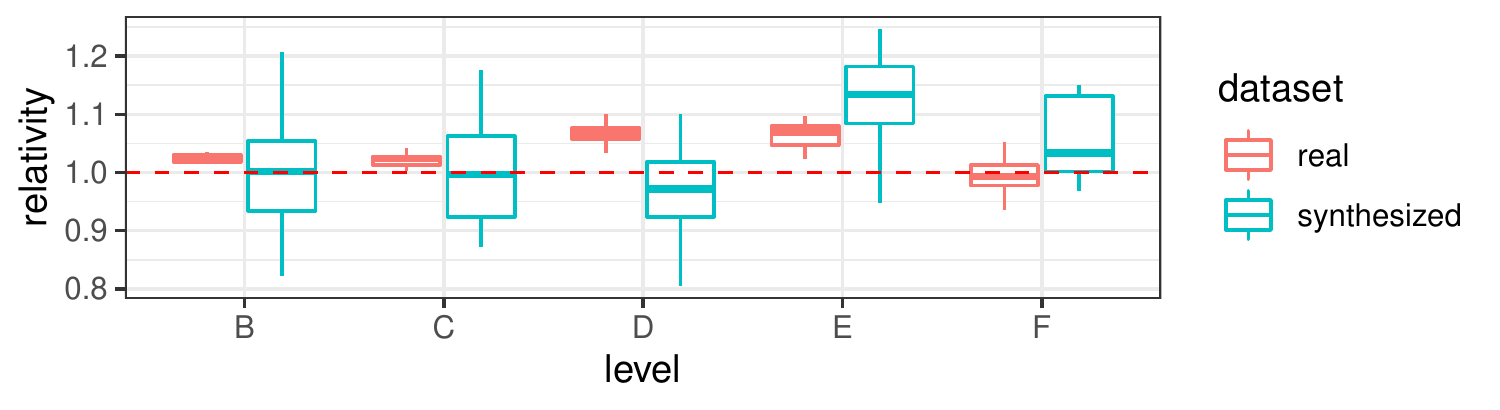}
\caption{\label{fig:relativities-area}Relativities of the Area variable across cross-validated models.}
\end{figure}

\begin{figure}
\centering
\includegraphics{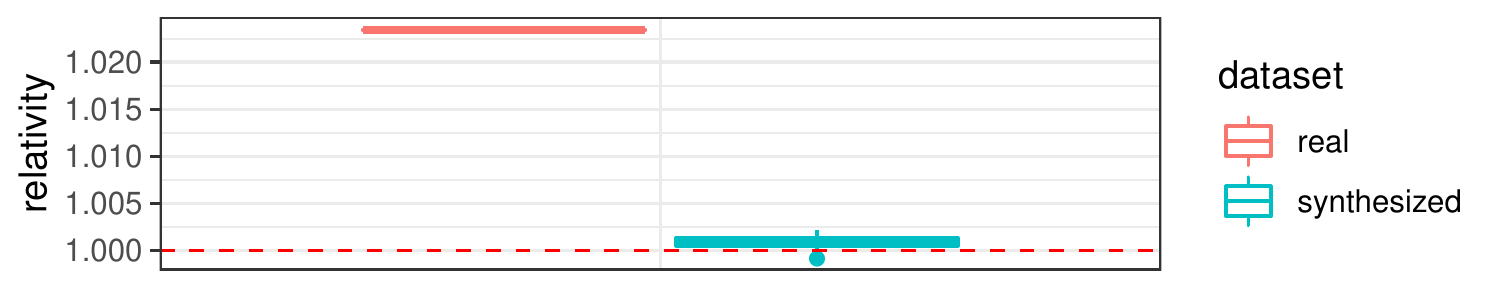}
\caption{\label{fig:relativities-bonus-malus}Relativities of the Bonus/Malus variable across cross-validated models.}
\end{figure}

\begin{figure}
\centering
\includegraphics{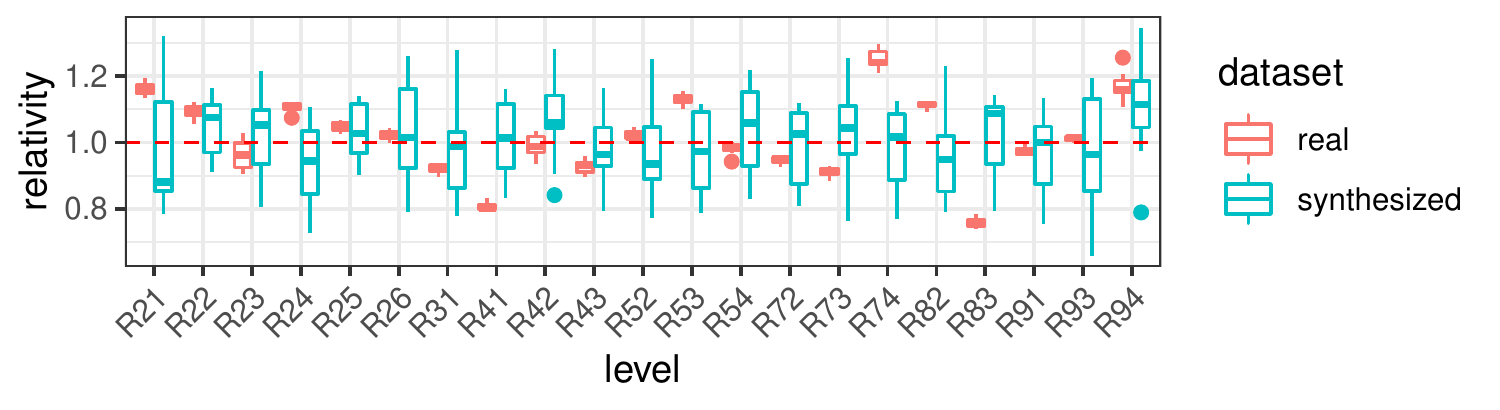}
\caption{\label{fig:relativities-region}Relativities of the Region variable across cross-validated models.}
\end{figure}

In terms of model relativities, we see a divergence of patterns between the synthetic
and real datasets. In Figures \ref{fig:relativities-area}, \ref{fig:relativities-bonus-malus}, and \ref{fig:relativities-region},
we plot the
relativities of select variables from
models trained during 10-fold cross-validation. We see that the trends are noticeably
different and that the variances of the relativities resulting from models fitted
on the synthetic datasets are larger than those resulting from models fitted on the
corresponding folds of the real dataset.

These explorations, which we recommend users perform on their datasets of interest,
enable them to form an initial assessment on whether the workflow
is appropriate for their use cases. For example, in our case, due to the instability
of GLM parameters, it may not be advisable to develop a rating plan using the synthesized
data. However, it may be appropriate to use the synthesized dataset to demonstrate
a procedure for developing such a plan.

We note that we do not perform hyperparameter tuning of CTGAN in these experiments,
so results could potentially be improved if dataset-specific tuning is performed.

\hypertarget{workflow}{%
\section{Workflow and Privacy}\label{workflow}}

To encourage usage, we implement an R interface to the original Python implementation
in the form of an open source package, \textbf{ctgan}.\footnote{\url{https://github.com/kasaai/ctgan}} We envisage this package to appeal to a
variety of personas in the field. As examples, companies who are looking to crowd-source
insights into their data gain an option for data disclosure, and researchers working
with confidential data can release reproducible workflows along with a sample dataset.
Depending on the use case, one can decide whether to share a single
sample dataset, or release a trained synthesizer, containing the generator only,
for users to generate data on demand.
In the latter case, one can further choose whether to make available the generator's
parameters, or to expose the generator through a Web service, while keeping the
model parameters private.

Assuming one has identified the dataset
and performed necessary anonymization and pre-processing, an example workflow
for data disclosure is as follows:

\begin{enumerate}
\def\labelenumi{\arabic{enumi}.}
\tightlist
\item
  Train a synthesizer using ctgan
\item
  Sample a dataset using the synthesizer
\item
  Perform post-processing on the generated dataset
\item
  Share the data
\end{enumerate}

In the case where one wishes to release the synthesizer itself, the steps
would resemble the following:

\begin{enumerate}
\def\labelenumi{\arabic{enumi}.}
\tightlist
\item
  Train a synthesizer using ctgan
\item
  Save the synthesizer model files
\item
  One of

  \begin{enumerate}
  \def\labelenumii{\alph{enumii}.}
  \tightlist
  \item
    Share the models files, along with post-processing code
  \item
    Provide access to a Web service that returns generated and post-processed datasets
  \end{enumerate}
\end{enumerate}

The mode of data disclosure depends on the goals of the data owner and the degree
to which the training data is already anonymized. The threat model
we are interested in is that of membership inference attacks. Specifically,
consider the case where an adversary has actual, but incomplete, information
of an individual claim (e.g., a subset of claim characteristic variables), and
also access to the data synthesizer. Would this adversary be able to reconstruct,
with high confidence, the full information on that specific claim? If so, one may
consider the event a privacy breach.

Data privacy with respect to GAN is an active area of research. \citet{chen2019gan} proposes
a taxonomy of different extents of model disclosure and perform experiments
on the efficacies of membership inference attacks on various architectures. \citet{hayes2017logan}
and \citet{hilprecht2019reconstruction} propose attacks on GANs for images to attempt
to recover images used in training. Defenses to these attacks have also been proposed,
such as PATE-GAN \citep{yoon2018pategan} and DPGAN \citep{xie2018differentially}, by making
the discriminator differentially private \citep{dwork2014algorithmic}.

While CTGAN, as implemented, is not differentially private, it can be extended to
be so. We remark that, even without formal privacy guarantees,
one can still avoid exposing personally identifiable information by removing
unnecessary data, such as social security numbers and birth dates of claimants, before
training. For an impactful attack to be theoretically possible, the adversary would
need to obtain information on an individual through other means, and have access
to the generator---either the model parameters or a sufficient number of queries
to the Web service. The
generator would also have to have been trained on granular enough data. In many
cases, publishing a sampled dataset may suffice for the goals of the data provider; without the
generator, vulnerabilities are minimized, and the provider can inspect the dataset
prior to release to further ensure nothing confidential is exposed.

\hypertarget{conclusion}{%
\section{Conclusion}\label{conclusion}}

In this paper, we adopt a recent methodology, CTGAN, for synthesizing tabular data, to
insurance datasets. We show that, with appropriate modifications, datasets
generated using this methodology can achieve high ML efficacy on representative insurance
datasets. To promote adoption within the insurance industry, we implement an
open source R interface to utilize the technique.

While this paper focuses on generating datasets with scalar values, future
work may include using generative modeling techniques to synthesize sequential data.
Sequential data generation is particularly interesting for claims reserving research,
where the instances of interest are cash flows over time. Related to data synthesis
are privacy requirements in data disclosure. On this subject, interesting questions
include what degree of privacy is needed for various types of data, and whether
differential privacy is warranted in different scenarios.

\bibliography{manuscript}

\end{document}